\documentclass{PoS}
\newcommand{\ee}{e^{+}e^{-}}
\newcommand{\leplep}{\ell^{+}\ell^{-}}
\newcommand{\jp}{J/\psi}

\newcommand{\psip}{\psi^{\prime}}

\newcommand{\pipi}{\pi^{+}\pi^{-}}

\newcommand{\fb}{{\rm fb}^{-1}}
\newcommand{\ra}{\rightarrow}
\newcommand{\rt}{\rightarrow}
\newcommand{\etal}{{\em et al. }}

\title{Spectroscopy results from Belle}

\ShortTitle{HQL 2010}

\author{\speaker{Sookyung Choi (for the Belle Collaboration)}
\\
        Gyeongsang National University\\
        E-mail: \email{schoi@gsnu.ac.kr}}

\abstract{
We report recent results on the charmonium and charmoniumlike states 
 based on a large data sample recorded at  the $\Upsilon(4S)$ 
and $\Upsilon(5S)$ resonances with the  Belle detector 
at the KEKB asymmetric-energy $e^+e^-$ collider.
}

\FullConference{The Xth Nicola Cabibbo International Conference on Heavy Quarks and Leptons,\\
		October 11-15, 2010\\
		Frascati (Rome) Italy}

\begin{document}

\section{Introduction}

Starting with the observation of the $\eta_{c}(2S)$ state,
a number of new states have been observed by Belle.
Some of these have been identified as being the predicted, but 
not-yet-seen, 
charmonium states, while
others, designated by $X,~Y~\&~Z$, are considered to be candidates 
for new types of charmoniumlike states such as
hybrid $c\bar{c}$-gluon states or 
 multiquark states either of the molecular type ($c\bar{q}\bar{c} q$) 
or the diquark-diantiquark~($cq\overline{cq}$) type.
Here, recent $XYZ$ state-related measurements are reported and examined  
in the context of possible charmonium assignments.

\section{The $X(3872)$}
The $X(3872)$ was discovered by Belle in the 
$\jp\pi^+\pi^-$ mass spectrum in exclusive $B \rightarrow K X(3872) $ 
decay~\cite{x3872-ref}
in a 140 $\fb$ data sample; it
was subsequently seen in three other experiments.
One curious fact about the $X(3872)$
is the near equality of its mass and the 
$m_{D^0}$+$m_{D^{*0}}$ threshold. 
The latest PDG world averaged mass~\cite{PDG} is $3871.56 \pm0.22$~MeV, 
while the $D^0 D^{*0}$ mass threshold is $3871.79 \pm 0.30$~MeV.
The CDF group reported that only the $J^{PC}$ options 
$1^{++}$ and $2^{-+} $ are compatible with the $\jp$ and $\pi\pi$ helicity angle 
distributions~\cite{cdf-JPC}.
It is also now well established that the $\pi\pi$ system in $X\ra \jp \pi^+\pi^-$
comes from $\rho\rightarrow \pi\pi$ decays.
%
%
\subsection{$ X(3872) \rt \gamma \jp (\psip)$ }

The observation of $X(3872)$ in $\gamma\jp (\psip)$ final states 
ensures that the charge-conjugation parity of the $X(3872)$ is $C=+1$.
The first reported evidence for $X(3872) \rt \gamma \jp$ was
given by Belle~\cite{Belle-omega} using 
in $B \rt K X(3872)$ decays in a 256$\fb$ data sample 
with a signal significance of $\sim 4\sigma$.

BaBar also reported $X(3872)$ decays to both $\gamma\jp$ and $\gamma\psip$
final states in the $B^+ \ra K^{+} \gamma \jp (\psip)$ process~\cite{BaBar-gammapsi},
with signal significances of 3.6~$\sigma$ and 3.5~$\sigma$, respectively.
Their measured product of branching fractions are
${\it B} (B^{+} \ra K^{+} X(3872)) \times {\it B} (X(3872) \ra \gamma J/\psi)
= (2.8 \pm 0.8) \times 10^{-6}$ and 
${\it B} (B^{+} \ra K^{+} X(3872)) \times {\it B} (X(3872) \ra \gamma \psip)
= (9.5 \pm 2.8) \times 10^{-6}$.
The ratio of these branching fractions is $\frac {{\it B} (X_{3872}\ra 
\gamma \psip)}
{{\it B} (X_{3872}\ra \gamma J\psi)} = 3.4\pm 1.4$, which is  
large for a $D\overline{D}^{*0}$ molecular state~\cite{Swanson} 

This year Belle studied the $\gamma J/\psi (\psip)$ final states
using their full data sample of $772 \times 10^{6} B\overline{B}$ events~\cite{gammapsi}.
The $B \rt K \gamma \jp$ channel is dominated by 
$B \rt K \chi_{c1}$; $\chi_{c1} \rt \gamma \jp$ decays and this is used 
as a calibration reaction; the branching fraction for
the well known $B^+ \ra K^{+}\chi_{c1}$ 
decay is measured to be $(4.94 \pm 0.35)\times 10^{-4}$,
which agrees well with the PDG value~\cite{PDG}.
Belle also reported first evidence for  $B\ra K\chi_{c2}$
(via $\chi_{c2} \ra \gamma\jp$) with
$3.6~\sigma$ significance. The branching fraction for 
$B^+ \ra K^{+} \chi_{c2}$ is measured to be $(1.11 \pm 0.37)\times 10^{-5}$.
The ratio of branching fractions 
$\frac {{\rm B} (B^{+}\ra K^{+} \chi_{c2})}
{{\rm B} (B^{+}\ra K^{+} \chi_{c1})} $ is $0.022 \pm 0.007$,
which is a measure of the factorization suppression factor for $J^{PC}=2^{++}$.
\begin{figure}
\includegraphics[width=0.65\textwidth]{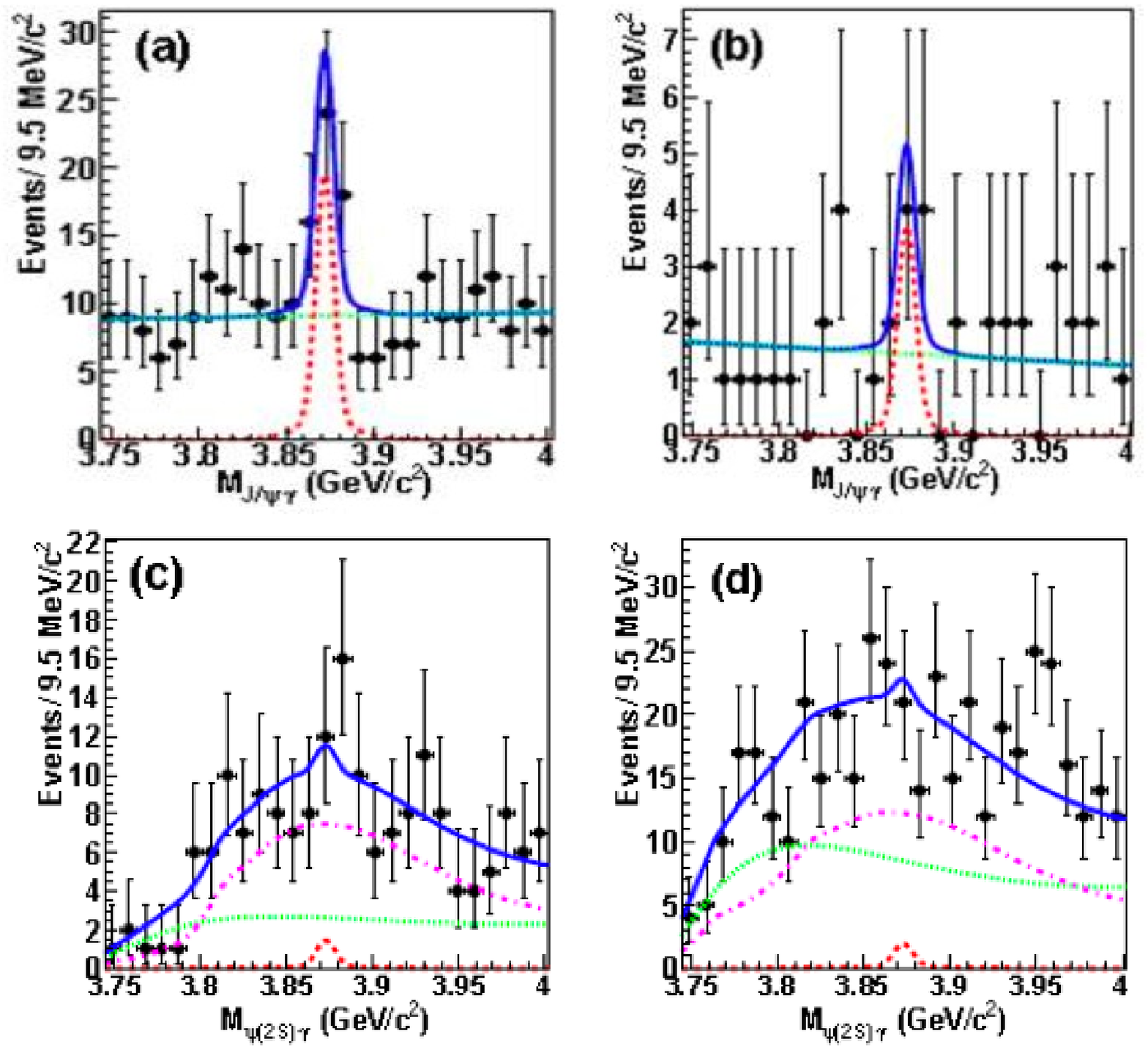}
\centering
\caption{
$\gamma\jp$ mass distributions (top) for (a) $B^{+}\rt K^{+} X(3872) $
and (b) $B^{0}\rt K^{0}_{s} X(3872) $ candidate events with subsequent 
$X(3872)\rt \gamma\jp$  decay.
$\gamma \psip$ mass distributions (bottom) for $B^{+}\rt K^{+} X(3872)$ 
candidate events with subsequent
 $X(3872)\rt \gamma\psi^{\prime}$ decay, where $\psip \rt \leplep$ in (c) 
and $ \pi^{+}\pi^{-} \jp$ in (d).
}
\label{gammapsi}
\end{figure}
In the same $\gamma\jp$ final state but at higher masses,
there is a clear $X(3872)\ra\gamma\jp$ signal with $4.9~\sigma$ 
significance.
Figure~\ref{gammapsi} shows $M_{\gamma\jp}$ mass distributions in
exclusive (a) $B^{+}\ra K^{+}\gamma \jp$ and (b) $B^{0}\ra K^{0}_{S}\gamma \jp$
decay. The product of branching fractions for $B\ra K X$ and $X\ra \gamma\jp$
is measured to be $(1.8 \pm 0.5)\times 10^{-6}$, which agrees with
the BaBar result.
However, there is no significant signal for $B^{+}\ra K^{+}X$ and 
$X\ra \gamma\psip$ decay.
Figures \ref{gammapsi} show the $M_{\gamma\psip}$ mass distributions in 
$B^{+}\ra K^{+}\gamma \psip$ decay for (c) $\psip \ra \leplep$ and 
(d) $\psip \ra \jp \pi^+ \pi^-$ decays.
An upper limit of the product of branching fractions is determined to be 
${\it B} (B^+ \ra K^+ X) \times (X\ra \gamma\psip) < 3.4 \times 10^{-6}$.
The ratio of branching fractions of its 90\% CL upper limit is
$\frac {{\it B} (X\ra \gamma\psip) }
{{\it B} (X\ra \gamma \jp)} < 2.1$, in
contradiction with the BaBar result.

\subsection{$X(3872) \ra \omega \jp $ }

In the $D^{0}\overline{D}^{*0}$ molecular model of Swanson~\cite{Swanson}, the
$J^{PC}$ is assumed to be $1^{++}$, in which case the $D^{0}\overline{D}^{*0}$
component is dominant,
with small admixtures of $\omega \jp$ and $\rho\jp$.  In this model,
$X(3872)\ra \pi^+ \pi^- \pi^0 \jp$ decays were predicted to occur 
at about half the rate for $X(3872) \ra \pi^+ \pi^- \jp$ decay. 
Belle performed a search for this $3\pi\jp$ decay mode. 

\begin{figure}
\includegraphics[width=0.55\textwidth]{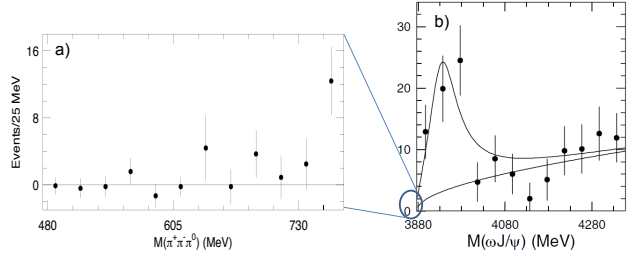} 
\centering
\caption{
a) The B-meson signal yields versus $3\pi$ invariant mass
in the $X(3872)$ region of $M_{X} -3\sigma< M_{X}< M_{X} + 1\sigma)$
for $B\rt K \pi^+ \pi^- \pi^0 \jp $ decay. b) $B \rt K \omega \jp$ 
signal yields versus $M(\omega \jp) $
}
\label{omega-belle}
\end{figure}

Figure~\ref{omega-belle}(b) shows the $Y(3940)$~\cite{y3940-belle} 
seen in $\omega\jp$ mass distribution in $B\rt K\omega\jp$ decay. 
In the $X(3872)$ mass 
region, which is at the right edge of the  kinematic boundary,
Belle observes a signal in the 
$3\pi$ mass spectrum corresponding to the sub-threshold decay
$X(3872)\rt \omega\jp$.
The measured number of signal events were $12.4\pm 4.1$, from which 
the ratio of the branching fractions  
$\frac{B(X_{3872} \rt \omega\jp)}{B(X_{3872} \rt \pi^+ \pi^- \jp)}$
is determined to be $1.0 \pm 0.5$~\cite{Belle-omega}.

This year BaBar reanalyzed $B\ra K \omega \jp$ final states
using a relaxed omega mass selection~\cite{BaBar-omega}
$0.5<m_{3\pi}<0.9 $~GeV and saw a similar $\omega$ signal.  Using their  
reported branching fraction, we obtain 
the combined ratio from Belle and BaBar to be
$\frac{B(X_{3872} \rt \omega\jp)}{B(X_{3872} \rt \pi^+ \pi^- \jp)}$
= $ 0.8\pm 0.3$.
In addition, BaBar reports that the $M_{3\pi}$ mass spectrum
from the $X_{3872} \rt J/\psi \omega$ final states is
suppressed near its  upper kinematic boundary
by a centrifugal barrier factor that is consistent with a $P$-wave.
Their $P$-wave ($2^-$) fit ($\chi^2$/NDF =3.53/5 )  to the $M_{3\pi}$ mass
distribution is favored over their
$S$-wave ($1^+$) fit ($\chi^2$/NDF =10.17/5). This would be bad for
a molecular interpretation of $X(3872)$, however, this corresponds
to only about a 1.5 $\sigma$ effect.
\subsection{Charmonium possibilities of $X(3872)$ }

From the CDF angular analysis results~\cite{cdf-JPC}, the only
two possible $J^{PC}$ assignments for the $X$ are $1^{++}$ and $2^{-+}$; 
all other $J^{PC}$ values are ruled out with high confidence.
In this section, we survey charmonium possibilities
for $X(3872)$ with these two $J^{PC}$ assignments.

For the $1^{++}$ assignment, the possible undiscovered charmonium state
is $\chi^{\prime}_{c1}$. For the $X(3872)=\chi^{\prime}_{c1}$
assignment, the following  puzzling questions arise.
\begin{itemize}
\item 
Since the mass of $\chi^{\prime}_{c2}$ is 
now known to be 3930~MeV~\cite{y-3}, the mass of 
$\chi^{\prime}_{c1}$ is expected to be $\sim 3905$ MeV. 
Therefore, the mass of $X(3872)$ is too low for it to 
be the $\chi^{\prime}_{c1}$.
\item 
Barnes \etal\cite{barnes-2005} estimated the partial width
for $\Gamma(\chi_{c1}\prime \ra \gamma \psip ) \sim 180$~keV,
while 
$\Gamma(\chi_{c1}\prime \ra \gamma J/\psi) \sim 14$~keV. 
So, the ratio of of partial widths $\frac{\Gamma(\chi_{c1}\prime \ra \gamma \psip)}
{\Gamma(\chi_{c1}\prime \ra \gamma J/\psi)}$ 
should be much bigger than unity. 
Therefore, the $\chi_{c1}\prime$ assignment would be a possible option 
if the BaBar measurement is right, which gives a large partial width 
for the $\gamma \psip$ mode.
\item 
$ \Gamma (X \rt \pipi \jp)=(3.4 \pm 1.2) \times \Gamma (X \rt \gamma \jp)$
is estimated to be about 45~keV using the Barnes 
value for $\Gamma_{\gamma \jp}$ 
and the measured ratio of $\frac{\Gamma_{\pi\pi J/\psi}}{\Gamma_{\gamma\jp}}$.
This partial width for $X \ra \pipi \jp$ decay is huge for 
an isospin-violating transition; other isospin 
violating transitions in the charmonium system 
are less than 1~keV
({\it e.g.}, $\Gamma(\psi' \rt \pi^{0} \jp)\sim 0.4$~keV~\cite{PDG}).
\end{itemize}

For the $2^{-+}$ assignment, the possible undiscovered charmonium state is 
the singlet $D$-wave state known as the $\eta_{c2} (^1D_2)$. 
For this assignment, the following questions arise.
\begin{itemize}
\item 
Y. Jia \etal\cite{YJia} estimated the partial widths
to be 
$\Gamma (\eta_{c2} \rt \gamma \psi ')\sim 0.4$ keV and 
$\Gamma (\eta_{c2} \rt \gamma \jp)\sim 9$ keV.
This favors the Belle measurement where the ratio 
$\frac{\Gamma (\eta_{c2} \rt \gamma \psi')}{\Gamma (\eta_{c2} \rt \gamma \jp)}$
is smaller.
\item
Using the well established mass of its multiplet partner, $M(\psi'') = 
3770$~MeV, the mass of the $\eta_{c2}$ is estimated to be 3837MeV. 
Thus, the $X(3872)$ mass is high for $\eta_{c2}$.
\item 
$\Gamma (X \rt \pipi \jp)=(3.4 \pm 1.2) \time \Gamma (X \rt \gamma \jp)$
is about 30~keV using the Jia $\gamma\jp$  width. This is also large for 
an isospin-violating transition.
\item 
For the $2^{-+}(\eta_{c2})$ assignment, the 
branching fraction for the $B^{+} \rt K^{+} \eta_{c2}$
is too high for a non-factorizable decay.
Other two $B^{+} \rt K^{+} h_{c}$ and $B^{+} \rt K^{+} \chi_{c2}$ decay, 
which is also non-factorizable and suppressed by an
angular momentum barrier,
are just barely  seen in the huge Belle data sample.
\item
The branching fraction for $\eta_{c2}\rt D \overline{D}^{*}$ is expected 
to be small~\cite{1d2}, but the averaged ratio from 
both Belle and BaBar is 
$\frac{\Gamma(X\rt DD*)}{\Gamma(X\rt \pi\pi\jp)}=9.5 \pm 3.1$, 
which is high for the $\eta_{c2}$.
\end{itemize}

\section{More $X$ and $Y$ states near 3940 MeV} 

Belle observed  three states near 3940~MeV  
via three different production and decay 
channels~\cite{y3940-belle,y-3,y-1}.
Among these three, the $Z(3930)$ state, which is produced 
in the $\gamma\gamma \rightarrow D\overline{D}$ process, 
is generally considered to be the charmonium $\chi_{c2}'$ state, even 
though the mass M=$3929\pm5\pm2$~MeV is somewhat
lower than potential model predictions. 
The $X(3940)$ is observed
in the $D \overline{D}^{*}$ mass spectrum from double charmonium 
production in $\ee \rightarrow J/\psi D^*\overline{D}$ annihilation and
the $Y(3940)$ is observed in the $\omega\jp$ mass
spectrum in $B\rightarrow K\omega\jp$ decays.
The mass and width of the $X(3940)~(Y(3940))$ are measured to be
M=$3942^{+7} _{-6} \pm 6$ ($3943\pm 11 \pm 13$)~MeV and 
$\Gamma=37^{+26} _{-15} \pm8$ ($87\pm 22 \pm 26 $)~MeV. 
Although the masses are similar, 
the $X(3940)$ and $Y(3940)$ appear to be
different states: the $X(3940)~(Y(3940))$ has not been
seen in the $\omega \jp$ ($D\overline{D}^*)$ final state
in $B\rightarrow X(Y)K$ decays.
%
%

It is important to search for 
$ \omega\jp ~($ or $D\overline{D}^*) $ in two-photon 
collisions, where its spin-parity of resonance 
is preferentially constrained to be $J^{p}$=$0^{\pm}$ or $2^{\pm}$.  
Belle observed a 
7.7~$\sigma$ enhancement
in the $\omega J/\psi$ system~\cite{uehara-omegajpsi} 
produced in the $\gamma\gamma \rightarrow \omega \jp$ process;
the mass and total width 
are measured to be M=$3915\pm3\pm2$~MeV and $\Gamma$=$17\pm 10\pm 3 $~MeV. 
This state, denoted by $X(3915)$, is probably related to one of the
three above-mentioned states in the 3.90-3.95 GeV mass region.
If we assume the $X(3915)$ is $0^+$ ($2^+$) resonance, the product of the 
two-photon decay width and branching fraction to $\omega\jp$ is 
determined to be
$\Gamma_{\gamma\gamma}(X(3915))\rm{B}(X\rightarrow\omega\jp)$=
$61\pm17\pm8 (18\pm5\pm2)$~eV for $J^p = 0^+ (2^+)$.  For comparison,
the measured product of the two-photon decay width and branching fraction 
for $Z(3930)\rightarrow D\overline{D}$ is $180\pm50\pm30 $~eV. 
If the $X(3915)$ is the $Z(3930)$~($\chi_{c2}^{\prime}$), the  ratio 
of branching fractions $\frac{\rm{BF}(\chi_{c2}' \rightarrow\omega\jp)} 
{\rm{BF}(\chi_{c2}' \rightarrow D\overline{D})}$ 
is large for an above-open-charm-threshhold charmonium state.
Also, for both the $0^+$ and $2^+$ options, if we assume
that the $\gamma\gamma$ 
partial width is $\sim$1~keV, which is typical for charmonium states,
$\Gamma_{\omega\jp}$ would be of the order of 1~MeV, which is
large for charmonium.

The $\Gamma (Y_{3940}\rt \omega \jp)$ partial width is also estimated 
to be large using the averaged product branching fraction 
from Belle and BaBar to be 
$B(B^+ \rt K^+ Y_{3940}) \times B(Y_{3940} \rt \omega \jp) = (5.0 \pm 0.8) 
\times 10^{-5}$ and the PDG averaged $\Gamma (Y_{3940})=40 ^{+18} _{-13}$~MeV. 
If we assume the maximum possible branching fraction for 
$B\rt K Y_{3940}$ is $10 \times 10^{-4}$ (the 
branching fraction for $B\rt K \jp$),  the partial width 
for $\Gamma( Y_{3940}\rt \omega\jp)$ is determined to be larger
 than order 1~MeV, which is 
large for  conventional charmonium.
%
%
%
%
%

\section{The $Y(4260)$ and $Y_{b}$ }

The $1^{--}$ $Y(4260)$ state was first discovered 
by BaBar~\cite{y4260-babar} and confirmed by Belle~\cite{y4260-belle} 
in the $\jp\pi^+\pi^-$ 
in radiative $\ee\rightarrow \gamma_{ISR}Y(4260)$ process.
The partial width for $Y \rt \pi^{+} \pi^{-} \jp$ 
is determined to be larger than 
0.5~MeV at the 90\% CL level by Mo 
\etal~\cite{XHMo}, which is 
much 
larger than that for 
$\psi^{\prime}\rt \pi^{+} \pi^{-} \jp$.
This large partial width is one of the remarkable properties of  the $Y(4260)$
that have led to various exotic interpretations of its quark content.
An interesting question 
is whether or not there exist 
counterparts in  the $s\overline{s}$ and/or $b\overline{b}$ quark systems.

\begin{figure}
\includegraphics[width=0.4\textwidth]{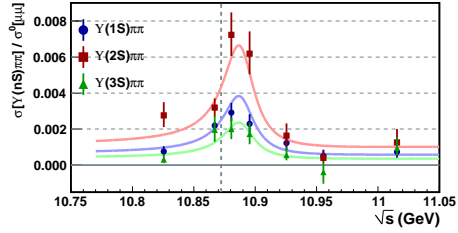}
\centering
\caption{The $\Upsilon(nS)\pi\pi$ (n=1,2 and 3) 
cross sections near the $\Upsilon(5S)$ resonance show peak structure
that deviates from the peak (vertical line) of $\Upsilon(5S)$ obtained 
from the hadronic cross sections.}
\label{up5s}
\end{figure}

Belle reported  an anomalously large 
$e^+e^-\ra \Upsilon(1,2S)\pi^+\pi^-$ production cross section  
near the peak of the $\Upsilon(5S)$ resonance at $\sqrt{s}\sim 10.87$ GeV
measured with a 21.7$\fb$ data sample~\cite{up5s}.
If they assume that the signal events only come from 
decays of the $\Upsilon(5S)$  resonance,
their extracted partial widths are $\sim$300 times larger than those 
for corresponding transitions from the $\Upsilon(4S)$.
Recently, Belle measured the energy dependence of
the  $\ee \rightarrow \Upsilon(nS) \pi\pi$ 
(n=1,2,3) production cross section using data accumulated 
at seven different cm energy points near the $\Upsilon(5S)$ resonance.
A new common peak structure was
observed for all three $\ee \ra \Upsilon(1,2,3S) \pi^+ \pi^-$
cross sections.
A fit using Breit-Wigner resonance function with a common mass and width
to these peaks, shown in Fig.~\ref{up5s}, gives a
mass and width of
M=$10888.4^{+2.7} _{-2.6}\pm1.2$~MeV and 
$\Gamma=30.7^{+8.3}_{-7.0} \pm 3.1$~MeV\cite{up5s-scan},
which are not consistent with any known
$b\overline{b}$ state such as the $\Upsilon(10860)$.
This can be considered to be a candidate for a $Y_b$-type state in the 
$b\overline{b}$ system.
\section{Charged $Z^+$ states}
A charged charmonium-like state could not be a $c \overline{c}$ charmonium state;
its minimal quark structure would have to be a 
$c\overline{c} u\overline{d}$ tetraquark arrangement.
The charged $Z(4430)^{+}$ state was first observed
by Belle as a peak in the $\pi^{+}\psip$ mass distribution
in exclusive $B \rt K \pi^{+}\psip$
decays\cite{z4430-ref}.  It was confirmed by a 
subsequent reanalysis using a Dalitz plot formalism~\cite{dalitz-z-ref} 
that includes all possible intermediate $K\pi$ resonances. 
This Dalitz analysis method was first employed
in the observation of two other
charged $Z_{1}^+$ and $Z_{2}^+$ states that are seen to decay
to $\pi^+ \chi_{c1}$ final states in exclusive 
$B\rightarrow K\pi^+\chi_{c1}$ decays~\cite{z1z2-ref}.  
The Dalitz-plot analysis demonstrated that these $Z$ states 
are not produced by reflections 
from any known and possibly unknown resonances in the $K\pi$ channel. 
However, BaBar searched for the $Z(4430)^+$, but did not see a 
significant signal~\cite{BaBar-Z}.
%
%
%
%
\acknowledgments{
This work was supported by the Korea Research Foundation Grant
funded by the Korean Government (KRF-2008-313-C00177)}.

\end{document}